\documentclass{PoS}

\title{Theoretical overview of $b \to s$ hadronic decays}

\ShortTitle{Theoretical overview of $b \to s$ hadronic decays}

\author{\speaker{Sebastian J\"ager}%
\\
       University of Sussex, Department of Physics and Astronomy,
       Falmer, Brighton BN1 9QH, UK\\
       E-mail: \email{S.Jaeger@sussex.ac.uk}}

\abstract{A wealth of data on hadronic $b \to s$ transitions is available from
the $B$-factories and will be improved at the LHCb experiment
and possible future super-$B$-factories. I review the theory of these decays
as it pertains to the search for physics beyond the Standard Model and
various puzzles in the present data.
}

\FullConference{12th International Conference on
                 $B$-Physics at Hadron Machines \\
		 September 7-11, 2009\\
		 Heidelberg, Germany.}

\begin{document}

\section{Introduction}
Charmless hadronic $b \to s$ transitions are a rich source of
information about the physics of the weak and/or TeV scales.
Their sensitivity
to short-distance physics derives from
the CKM hierarchy and a GIM cancellation which combine to
suppress contributions at tree-level in the weak
interaction or through light-quark loops.
As a consequence, the Standard-Model (SM) amplitudes are governed by
the combination
\begin{equation}
  V_{ts}^* V_{tb} \times \frac{1}{16 \pi^2} \times \frac{m_B^2}{M_W^2} 
\sim 10^{-6} .
\end{equation}
The resulting rareness of these modes makes them sensitive to contributions of
new particles with TeV-scale masses,
so we should expect deviations from the
Standard Model. The task is to disentangle SM and
new-physics (NP) contributions in a given mode, such that a possible
NP signal can be recognized, and to identify
those observables, or combinations of them, where this is best possible.
More ambitiously, one may want to quantify a signal in terms of
NP-model parameters.

\begin{figure}
\centerline{
\includegraphics[width=.6\textwidth]{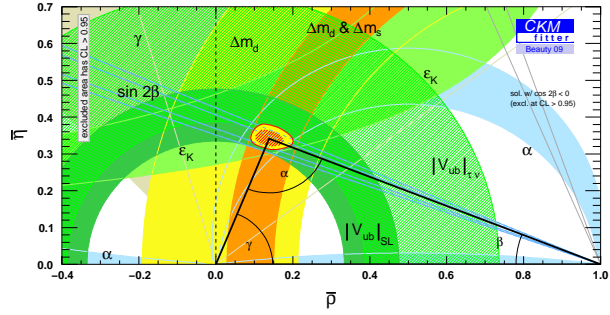}
}
\caption{Global CKM fit as shown at this conference \cite{ckmfitter}}
\label{fig:ckmfit}
\end{figure}

It is worth contrasting the $b \to s$ transitions with the $b \to d$
ones. Here, the CKM hierarchy is different, such that tree-level
contributions involving $V_{ub} \propto (\bar \rho - i \bar \eta)$
can compete with or dominate over
loop contributions involving $V_{td} \propto (1 - \bar \rho - i \bar \eta)$.
Indeed, $b\to d$  hadronic decays, together with $b \to u$
semileptonic and, by now, purely leptonic $B^+ \to \tau \nu_\tau$ decays,
provide the main input to the global CKM fit
(Figure \ref{fig:ckmfit}).
The two dominant inputs are the
ratio of $B_d$ and $B_s$ mass differences (orange ring) and the
mixing-induced CP violation in $B_d \to J/\psi K_S$ (blue wedge)
derive from the $B - \bar B$ mixing amlitudes,
which are again loop processes.
%(The short-distance
%physics cancels out in the Standard Model, but not necessarily
%beyond.)
Two constraints in a plane will generically intersect in
a discrete set of points, and the most
significant consistency check is through the ``$\alpha$''
measurements in $B_d \to {\pi \pi, \pi \rho, \rho \rho}$
transitions (shown as light blue ``half moon'' in the Figure).
Hence the consistency of the CKM fit
at present allows ${\cal O}(10 \%)$ NP
effects. Beyond this level, NP contributions to different observables
would have to conspire to maintain the observed level of agreement.\footnote{
The picture may change as progress in lattice QCD makes more
precise predictions for $B$ meson mixing, $B \to \tau \nu$, and
$\epsilon_K$ possible. Interestingly, a significantly improved
calculation of $B_K$ \cite{latticeepsk} indicates a 
tension with the aforementioned CKM determinations at about the
$2 \sigma$ level \cite{epsktension}.}
On the other hand, the combination $V_{ts}^* V_{tb}$ relevant to
$b \to s$ transitions
is very weakly dependent on $\bar \rho$ and $\bar \eta$. Hence
these processes are determined, in principle, with a small
parametric uncertainty in the SM. Moreover, the
consistency of the CKM fit has little to say about
new physics in $b \to s$ transitions.
Indeed, several puzzles have shown up in recent years in the data,
notably
\begin{enumerate}
\item  \label{it:Spar}
time-dependent CP violation in $b \to s$ decays of $B^0_d$ mesons to
a CP eigenstate. In the SM, one expects to measure
$-\eta_{CP} S \approx \sin 2 \beta$, but some modes show a deviation
(Figure \ref{fig:sin2betaeff}).
None of these is very significant at the moment, but this might change
with more precise data becoming available from LHCb and, eventually,
a super-$B$ factory.
\item
The time-dependent CP violation in $B_s \to J/\psi \phi$, in combination
with lifetime difference and semileptonic asymmetry,
determines the phase of the mixing amplitude to be \cite{bsmphase}
\begin{equation}
 \phi_{B_s} \in (-168, -102)^\circ \cup (-78, -11)^\circ   ,
\end{equation}
about $2.2 \sigma$ from the SM, with much better statistics ahead
at Tevatron and LHCb.
The theory is reviewed in a separate talk
at this conference \cite{balltalk}.
\item \label{it:ACP}
Direct CP asymmetries in $B \to \pi K$ decays. These modes have
received attention for several years. It has
been stressed that
$A_{\rm CP}(B^+ \to \pi^0 K^+) \not = A_{\rm CP}(B^0
\to \pi^- K^+)$ at $5 \sigma$ significance \cite{bellenature}.
The verdict is less clear, since the SM
does {\em not} predict identical asymmetries.
\end{enumerate}

\section{Hadronic decay amplitudes}

Interpreting items \ref{it:Spar} and \ref{it:ACP} requires 
knowledge about hadronic decay amplitudes, which always
involve nonperturbative QCD.
As the latter is generally under limited control,
approximations are necessary, either neglecting some small parameter or
expanding in it.

\begin{figure}
\centerline{
\includegraphics[width=.4\textwidth]{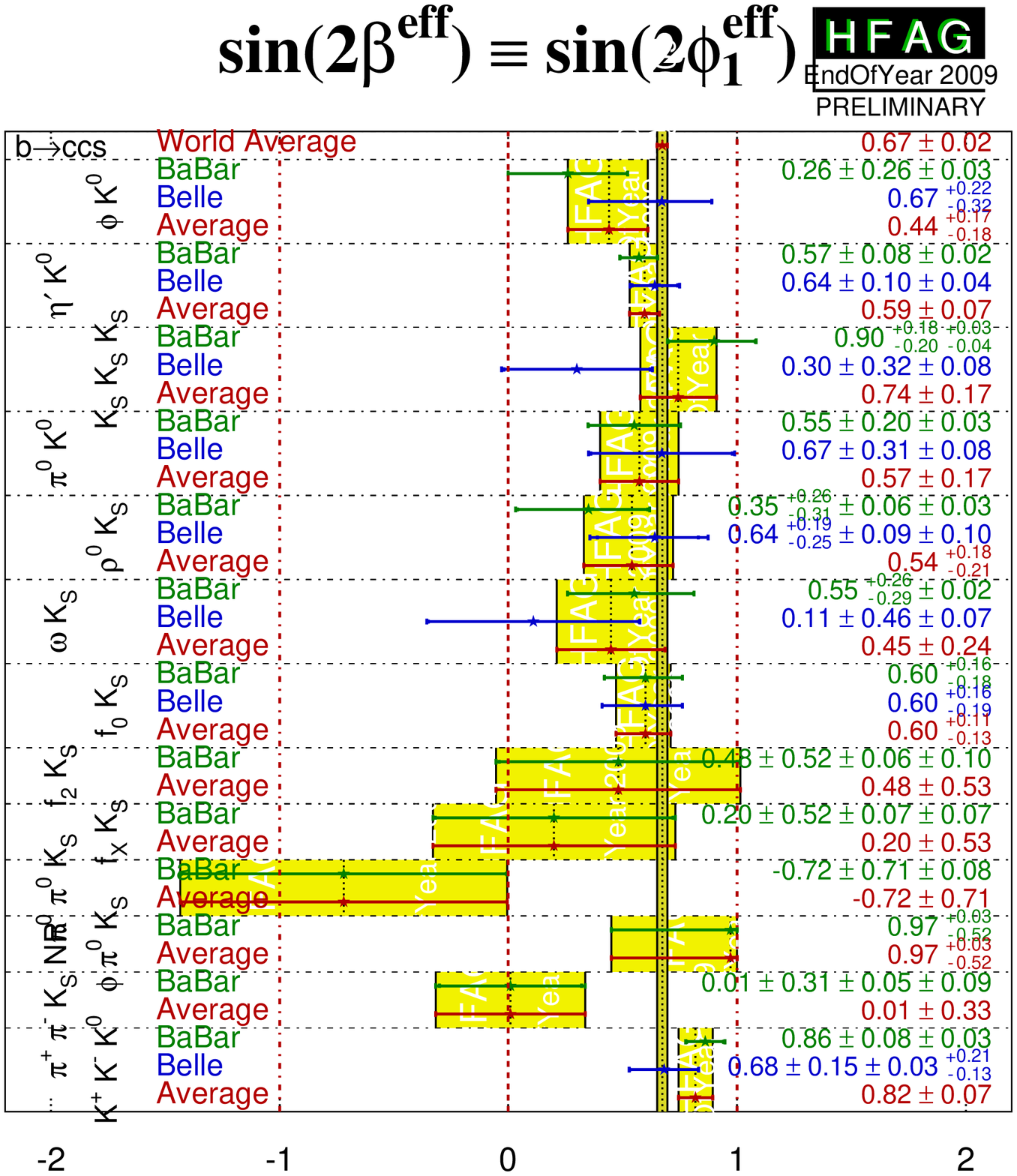}
\hskip5mm
\includegraphics[width=.43\textwidth]{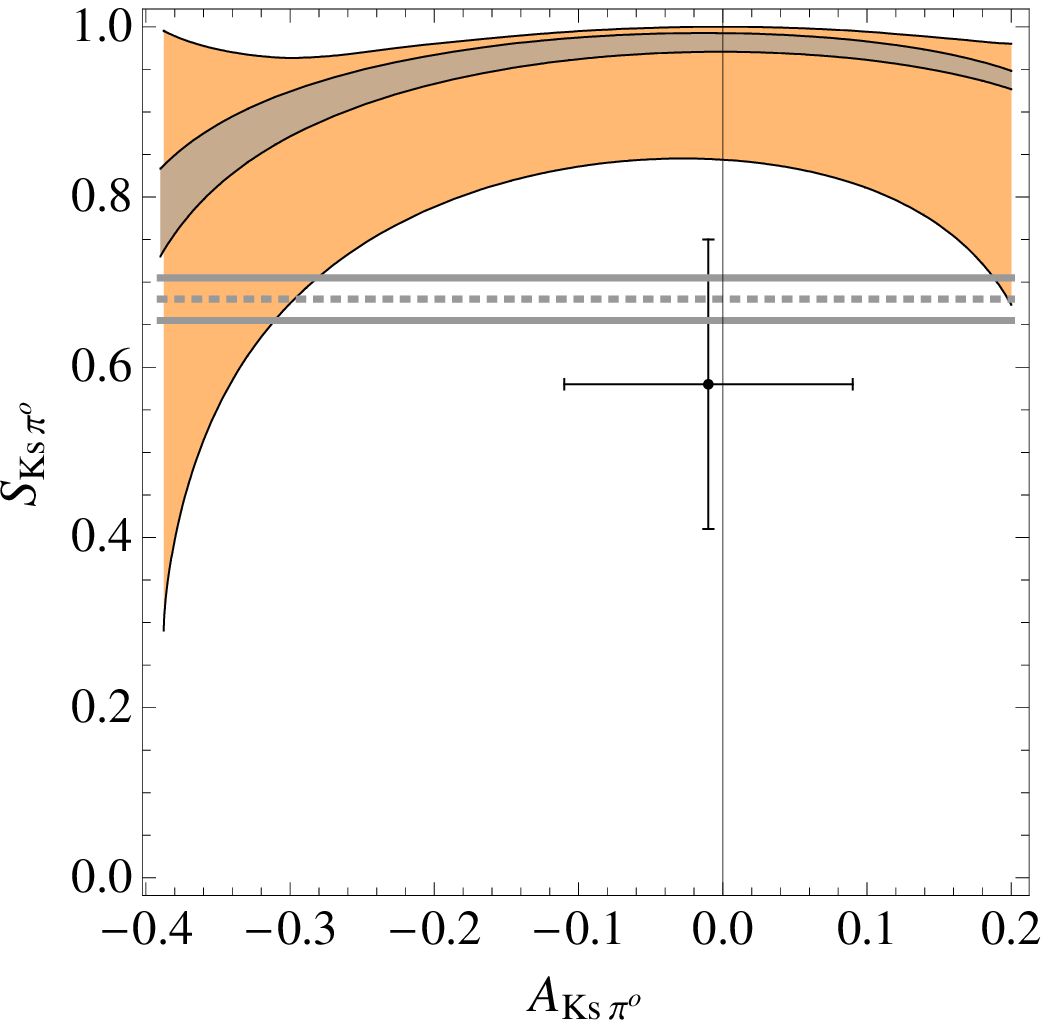}
}
\caption{Left:
Measurements of mixing-induced CP asymmetries in $b\to s$
penguin transitions as compiled by the HFAG \cite{Barberio:2008fa}.
Right: Constraints from decay rate data in the
$(A,S)$ plane for $B \to \pi^0 K_S$ ($A=-C$) \cite{Fleischer:2008wb}}
\label{fig:sin2betaeff}
\end{figure}

For any $b \to s$ transition to a final state $f$, we can write
\begin{eqnarray}
   {\cal A}_f &\equiv& {\cal A}(B \to f)
         = V_{us} V_{ub}^* T_f + V_{cs} V_{cb}^* P_f + P_f^{\rm NP} ,
         \\
   \bar {\cal A}_{\bar f} &\equiv& {\cal A}(\bar B \to \bar f)
         = V_{us}^* V_{ub} T_f + V_{cs}^* V_{cb} P_f + P_{\bar f}^{\rm NP} ,
\end{eqnarray}
where $T_f$ and $C_f$ are CP-even ``strong'' amplitudes and
$P_f^{\rm NP}$, $P_{\bar f}^{\rm NP}$ are new-physics contributions.
CKM unitarity has been used to eliminate the
combination $V_{ts} V_{tb}^*$ ($V_{ts}^* V_{tb}$).
%equivalently one may choose to eliminate $V_{cs} V_{cb}^*$ ($V_{cs}^* V_{cb}$),
%which implies $T_f \to T_f - P_f$, $P_f \to -P_f$ .
Branching fractions and CP asymmetries are
functions of the magnitudes and relative phases of the strong
amplitudes, as well as magnitudes and phases of the CKM elements.
For instance, if $f$ is a CP eigenstate,
$|\bar f \rangle = \eta_{\rm CP}(f) | f \rangle$, then the time-dependent
CP asymmetry is given as
\begin{equation}
  A_{\rm CP}(f; t) \equiv \frac{\Gamma(\bar B(t) \to f) -
    \Gamma(B(t)\to f) }{ \Gamma(\bar B(t) \to f) + \Gamma(B(t) \to f)}
 \equiv - C_f \cos \Delta m_d\, t + S_f \sin \Delta m_d\, t ,
\end{equation}
\begin{equation}
   C_f = \frac{1-|\xi|^2}{1+|\xi|^2}, \qquad
   S_f = \frac{2 {\rm Im} \xi}{1+|\xi|^2} , \qquad
   \xi = e^{-i 2 \beta} \frac{{\cal A}(\bar B
     \to f)}{{\cal A}(B \to f)}
       = -\eta_{\rm CP}(f) e^{-i 2 \beta} \frac{V_{cs}^* V_{cb} + \dots}{V_{cs}
         V_{cb}^* + \dots} .
\end{equation}
Here the dots are proportional to the ratio $T_f/P_f$, multiplied by
CKM factors of ${\cal O}(\lambda^2)$. If the tree amplitudes are
neglected, then
$- \eta_{\rm CP}(f) S_f = \sin(2 \beta)$ results to very good approximation.
While experimentally (Figure \ref{fig:sin2betaeff}) the various modes
are in reasonable agreement with each other and the
determination of $\sin 2\beta$ from $b \to c \bar c s$ transitions, the
suggestive pattern of the central values begs the question whether it
could be caused by the neglected SM tree amplitudes, or one has to
invoke NP terms $P_f^{\rm NP}$.
Quantitative information on the
amplitudes derives from (i) flavour-$SU(3)$ (and isospin)
relations \cite{su3}
together with measurements of $b\to d$ transitions and
(ii) the heavy-quark expansion in $\Lambda_{\rm QCD}/m_b$ (QCDF
\cite{Beneke:1999br} and its effective-field-theory formulation in
SCET \cite{scet,Williamson:2006hb,Beneke:2006mk},
and the somewhat different ``pQCD'' approach \cite{Keum:2000ph}).
Guidance on the relative importance of amplitudes follows from
(iii) Cabibbo counting and
(iv) the large-$N$ expansion \cite{largen}.
(i), (ii),
and (iv) involve the subdivision of the ``physical'' tree and
penguin amplitudes into several ``topological'' amplitudes,
\begin{eqnarray}
  T_{M_1 M_2} &=& \Big[ A_{M_1 M_2} (
\alpha_1(M_1 M_2) + \alpha_2(M_1 M_2) + \alpha_4^u(M_1 M_2) )
 \\ && \nonumber
+ B_{M_1 M_2} (b_1(M_1 M_2) + b_2(M_1 M_2) + b_3^u(M_1 M_2) +
b_4^u(M_1 M_2) ) 
+ {\cal O}(\alpha) \Big] \; + (M_1 \leftrightarrow M_2) \; ,
\\
  P_{M_1 M_1} &=& \Big[ A_{M_1 M_2} \alpha_4^c(M_1 M_2) 
+ B_{M_1 M_2} (b_3^c(M_1 M_2) + b_4^c(M_1 M_2) ) + {\cal O}(\alpha) \Big] 
 + (M_1 \leftrightarrow M_2) \; ,
\end{eqnarray}
where we employ the notation of \cite{Beneke:2003zv,Beneke:1999br},
which is general but is particularly suited for the heavy-quark expansion.
 $A_{M_1 M_2}$ and $B_{M_1 M_2}$ are normalization factors which by
 convention contain certain form factors and decay constants. 
The $\alpha_i$ and $b_i$ denote the different topological amplitudes.
Often, $A \alpha_1$ and $A \alpha_2$ are written as $T$ and $C$, $A
\alpha_4^c$ as $P_{ct}$, etc., or variations thereof. Table \ref{tab:topo}
summarizes the counting in the various small parameters.
\begin{table}[tb]
\centering
\caption{Hierarchies among topological amplitudes from expansions in
  the Cabibbo angle $\lambda$, in $1/N_c$, and in $\Lambda_{\rm QCD}/m_b$.
(Some amplitudes, such as electroweak penguins, are omitted from the list.)
\label{tab:topo}}
\begin{tabular}{c|cccccccccc}
& $\alpha_1$ & $\alpha_2$ & $\alpha_4^u$ & $\alpha_4^c$ &
$\alpha_{3EW}$ & $\alpha_{4EW}$ &
$b_3^c$ & $b_4^c$ & $b_1$ & $b_2$ \\
& $(T)$ & $(C)$ & $(P_{ut})$ & $(P_{ct})$ &
$(P_{\rm EW})$ & $(P_{\rm EW}^{\rm C})$ & & & $(E)$ & $(A)$ \\[2mm]
\hline
Cabibbo $(b\to d)$ & \multicolumn{10}{c}{all amplitudes are
${\cal O}(\lambda^3)$ } \\[1mm]
Cabibbo $(b\to s)$ & $\lambda^4$ & $\lambda^4$ &
$\lambda^4$ & $\lambda^2$ & $\lambda^2$ & $\lambda^2$ & $\lambda^2$ &
$\lambda^2$ & $\lambda^4$ & $\lambda^4$  \\
$1/N$ & $1$ & $\frac{1}{N}$ & $\frac{1}{N}$  & $\frac{1}{N}$ &
$1$ & $\frac{1}{N}$ & $\frac{1}{N}$ & $\frac{1}{N}$ & $\frac{1}{N}$ & $1$ \\ 
$\Lambda/m_b$ & $1$ & $1$ & $1$ & $1$ & $1$ & $1$ & $\Lambda/m_b$ & $\Lambda/m_b$ & $\Lambda/m_b$ & $\Lambda/m_b$ \\
\end{tabular}
\end{table}
At the quantitative level, the leading-power amplitudes $\alpha_1$,
$\alpha_2$, \dots can be factorized  \cite{Beneke:1999br} into
products of ``hard kernels'' that can be computed
order by order in perturbation theory and include all strong
(rescattering) phase information, and nonperturbative
normalization factors such as $f_+^{B \pi}(0) f_K$ or $f_B f_K f_\pi$
(usually factored out into $A_{M_1 M_2}$ and $B_{M_1 M_2}$). This
statement holds up to generally incalculable $\Lambda/m_b$
corrections. Certain amplitudes (annihilation amplitudes $b_i$) are
altogether power-suppressed and not calculable. See \cite{Antonelli:2009ws}
for more details.
Over the last years, a number of higher-order (NNLO) calculations of the
kernels have been performed \cite{factnnlo}.
The main phenomenological findings can be
summarized as follows.
\begin{itemize}
\item The colour-allowed trees $\alpha_1$ are well behaved in perturbation
theory, with overall uncertainties at the few-percent level (not
counting the nonperturbative normalization).
\item The colour-suppressed trees $\alpha_2$ show cancellations within the
(well-behaved) perturbative part, and  suffer from a large
uncertainty in the normalization and sensitivity to power
corrections. Attaching an ${\cal O}(1)$ uncertainty to the
(small) theoretical prediction using the power-correction model
of \cite{Beneke:1999br} would still
imply $|\alpha_2/\alpha_1|< {\cal O}(1/2)$.
\item The topological (QCD) penguin amplitudes are also under good
control (but only a subset of NNLO corrections is known),
but are phenomenologically indistinguishable from the incalculable
(formally power-suppressed) penguin annihilation amplitudes.
\item The colour-allowed and colour-suppressed electroweak
penguins amplitudes behave qualitatively like the colour-allowed
and colour-suppressed trees, respectively.
\end{itemize}
Further recent work focussing on phenomenological issues
can be found in \cite{factpheno}, and a new take on
long-distance charm penguins in \cite{Beneke:2009az}.

\section{Phenomenological applications}
\begin{table}[htb]
\centering
\caption{Predictions for $\Delta S$ defined in the text for several
penguin-dominated modes. {\em From \cite{Antonelli:2009ws}; see therein
for details, in particular the meaning and comparison of errors.}
\label{tab:DeltaS}}
\begin{tabular}{c||c|c|c|c}
mode & QCDF/BBNS \cite{Beneke:2005pu} & SCET/BPRS
\cite{Williamson:2006hb,Wang:2008rk} & pQCD \cite{Li:2006jv} &
experiment \cite{Barberio:2008fa} \\
\hline
$\phi K_S$ & $0.01$ \dots $0.05$ & $0$ / $0$  & $0.01$ \dots $0.03$ & $-0.23 \pm 0.18$ \\
$\omega K_S$ & $0.01$ \dots $0.21$ & $-0.25$ \dots $-0.14$ / $0.09$ \dots $0.13$ & $0.08$ \dots $0.18$ & $-0.22 \pm 0.24$ \\
$\rho^0 K_S$ & $-0.29$ \dots $0.02$ & $0.11$ \dots $0.20$ / $-0.16$ \dots $-0.11$ & $-0.25$ \dots $-0.09$  & $-0.13 \pm 0.20$ \\
$\eta K_S$ & $-1.67$ \dots $0.27$ & $-0.20$ \dots $0.13$ / $-0.07$ \dots $0.21$ &  &  \\
$\eta' K_S$ & $0.00$ \dots $0.03$ & $-0.06$ \dots $0.10$ / $-0.09$ \dots $0.11$ & & $-0.08 \pm 0.07$ \\
$\pi^0 K_S$ & $0.02$ \dots $0.15$ & $0.04$ \dots $0.10$  & & $-0.10 \pm 0.17$ \\
\end{tabular}
\end{table}
Several authors have estimated the tree ``pollution'' in the hadronic
$b\to s$ penguins combining experimental data and
heavy-quark-expansion calculations in different ways. Their results,
compared in Table \ref{tab:DeltaS}, are in general agreement with each
other (as they should) and can be compared to
Figure \ref{fig:sin2betaeff}. Clearly,
the SM does not produce the pattern of experimental
(central) values.
While the significance of the measured $\Delta S$ values is low for all
modes, in the case of $\pi^0 K_S$ one can perform a combined analysis
of all $B \to \pi K$ decay data to
get a somewhat stronger ``signal''.
The method discussed here \cite{Fleischer:2008wb}
(see also \cite{Gronau:2008gu}) invokes the well-known
isospin symmetry relation
\vskip-1mm
\begin{equation}
\sqrt{2} {\cal A}(B^0 \to \pi^0 K^0) + {\cal A}(B^0 \to \pi^- K^+)
   = - \Big[ (\hat T + \hat C) e^{i \gamma} + \hat P_{\rm ew} \Big]
\equiv 3 A_{3/2} .  
\end{equation}
\vskip-0.5mm
This relation, and a similar one for the CP conjugates,
allows to fix all four complex decay amplitudes from the four decay
rates {\em if} the isospin-$3/2$ amplitudes are known, up to a
four-fold ambiguity.
The latter can indeed be obtained as
\vskip-2mm
\begin{equation}
  3 A_{3/2} = - R_{T+C} |V_{us}/V_{ud}| \sqrt{2} |A(B^+\to\pi^+\pi^0)|
    \Big(e^{i \gamma} - 0.66 \frac{0.41}{R_b} R_q \Big) ,
\end{equation}
where $R_b$ is a side of the unitarity triangle and
$R_{T+C} = 1.23^{+0.02}_{-0.03}$ and $R_q =
(1.02^{+0.27}_{-0.22}) e^{i (0^{+1}_{-1})^\circ}$
quantify $SU(3)$ breaking, with uncertainties
obtained in a QCDF calculation.
Fixing the ambiguity by a (minimal) usage of either QCDF or $SU(3)$,
one obtains a prediction of
$S_{\pi^0 K_S}$ (Figure \ref{fig:sin2betaeff}) from the remaining data.
This is one of many ways of visualizing the tension in the $\pi K$
system, distinguished perhaps by a particularly limited use of
uncertain theoretical predictions or assumptions. A future
perspective on the uncertainty is also indicated (thin band).
For
more on NP in $B \to \pi K$,
see \cite{bpiknp}.

One can also attempt to compute directly the difference in direct
CP asymmetries. Unfortunately, this involves the uncertain
colour-suppressed tree amplitude, and the significance of this
discrepancy is currently difficult to quantify.
Making no assumptions about $C$, one still has the relation \cite{bpikiso}
$A_{\rm CP}(K^+\pi^-)+A_{\rm CP}(K^0\pi^+) \approx
  A_{\rm CP}(K^+\pi^0)+A_{\rm CP}(K^0 \pi^0)$,
which is satisfied by the current experimental data
\cite{Gronau:2008gu}, and expected to hold (in general) to few-percent level.

\section*{Acknowledgment}
I thank the organizers for
a superb conference experience within and outside the lecture theatre.

\end{document}